\documentclass[sigconf]{acmart}

\usepackage{colortbl}
\usepackage{multirow}
\usepackage{tabularx}
\usepackage{balance}
\usepackage{subcaption}
\usepackage{longtable}
\usepackage{supertabular}
\usepackage{tabularray}
\usepackage{tabularx}
\usepackage[font=small,labelfont=bf]{caption}

\usepackage[framemethod=tikz]{mdframed} %
\usepackage{enumitem}

\usetikzlibrary{calc}
\definecolor{labelColorBig}{RGB}{102, 102, 255}
\newcommand{\overviewLabelBig}[1]{\tikz[font=\sffamily, baseline={($ (current bounding box.center) - (0,.5em) $)}] 
\fill[fill=labelColorBig] (0,0) circle (0.5 em) node[text=white] {#1};}

\usepackage{soul}
\usepackage{adjustbox}
\usepackage{caption}
\usepackage{tikz}
\usepackage{xcolor}
\usepackage{refcount}
\usepackage{hyperref}

\usepackage[normalem]{ulem}
\usepackage{soul}

\newif\ifdraft
 \drafttrue 

\newcommand{\boldification}[1]{}

\usepackage[T1]{fontenc}
\usepackage{color, colortbl}
\definecolor{Gray}{gray}{0.9}
\newcolumntype{L}[1]{>{\raggedright\let\newline\\\arraybackslash\hspace{0pt}}m{#1}}
\newcolumntype{C}[1]{>{\centering\let\newline\\\arraybackslash\hspace{0pt}}m{#1}}
\newcolumntype{R}[1]{>{\raggedleft\let\newline\\\arraybackslash\hspace{0pt}}m{#1}}

\newif\ifdraft
\drafttrue 

\AtBeginDocument{%
  \providecommand\BibTeX{{%
    \normalfont B\kern-0.5em{\scshape i\kern-0.25em b}\kern-0.8em\TeX}}}

\begin{document}

\title{Unveiling Diversity: Empowering OSS Project Leaders with Community Diversity and Turnover Dashboards}

\author{Mariam Guizani}
\affiliation{%
  \institution{Oregon State University}
  \city{Corvallis}
  \state{Oregon}
  \country{USA}}
\email{guizanim@oregonstate.edu}

\author{Zixuan Feng}
\affiliation{%
  \institution{Oregon State University}
  \city{Corvallis}
  \state{Oregon}
  \country{USA}}
\email{fengzi@oregonstate.edu}

\author{Emily Judith Arteaga}
\affiliation{%
  \institution{Oregon State University}
  \city{Corvallis}
  \state{Oregon}
  \country{USA}}
\email{arteagae@oregonstate.edu }

\author{Luis Cañas-Díaz}
\affiliation{%
  \institution{Bitergia}
  \city{Madrid}
  \country{Spain}}
\email{lcanas@bitergia.com}

\author{Alexander Serebrenik}
\affiliation{%
  \institution{Eindhoven University of Technology}
  \city{Eindhoven}
  \country{Netherlands}}
\email{a.serebrenik@tue.nl}

\author{Anita Sarma}
\affiliation{%
  \institution{Oregon State University}
  \city{Corvallis}
  \state{Oregon}
  \country{USA}}
\email{Anita.Sarma@oregonstate.edu}

\begin{abstract}
Managing open-source software (OSS) projects requires managing communities of contributors. In particular, it is essential for project leaders to understand their community's diversity and turnover. 
We present CommunityTapestry, a dynamic real-time community dashboard, which presents key diversity and turnover signals that we identified from the literature and through participatory design sessions with stakeholders. 
We evaluated CommunityTapestry with an OSS project's contributors and Project Management Committee members, who explored the dashboard using their own project data. Our study results demonstrate that CommunityTapestry increased participants' awareness of their community composition and the diversity and turnover rates in the project. It helped them identify areas of improvement and gave them actionable information. 
\end{abstract}




\maketitle
\section{Introduction}
\label{sec:intro}
Diversity has been shown to have many positive effects on organizations, including a better ability to innovate, a better reputation as ethical citizens, and a better ``bottom line'' for businesses ~\cite{phillips2014diversity, page2019diversity, larkin2020diversity}. 

Research has investigated the influence of different diversity factors on productivity in Open Source Software (OSS). 
For instance, studies have demonstrated the impact of gender diversity on productivity~\cite{vasilescu2015gender} and the significance of organization affiliation diversity in ensuring project sustainability and success~\cite{yin2021sustainability}. Research also suggests that the concept of \emph{coopetition}, where competing companies use OSS to collaborate on shared challenges, leads to reduced workload, heightened productivity, and improved product quality~\cite{guizani2023rules}. 
On the other hand, a non-diverse community signals an unwelcoming environment to companies signifying a lack of neutral governance \cite{guizani2021long}, or to individuals signifying an non-inclusive environment \cite{guizani2022perceptions}. This can, in turn, result in a high turnover rate \cite{sharma2012examining, foucault2015impact}, where OSS projects struggle to attract and retain diverse contributors, especially women~\cite{terrell2017gender, lee2019floss}. 

Some project leaders recognize the positive effects of diversity, the adverse effects of lack thereof, and desire more diversity in their projects~\cite{qiu2023climate}. However, they struggle to find the time and bandwidth to monitor the state of diversity in their community and may miss important signals as explained by a Lead Program Manager for a Large OSS Company~\cite{germonprez2018eight}: \textit{``We try to determine the rough diversity of employment in our projects. The rough metric we have is about two-thirds of our contributors do not work for our primary corporate sponsor. That is extraordinarily critical when you look at both the objectives of our primary sponsor as well as the health of the community.''} 

In fact, keeping on top of everyday project tasks, such as pull-request reviews to ensure software quality, is already a challenge for project leaders~\cite{ferreira2021shut}. Project leaders report being stressed and feeling burnout, overwhelmed by the number of requests they receive (e.g., bug reports, support requests)~\cite{raman2020stress}. Hence, a more systematic support of monitoring diversity and implementing actions to improve it is necessary.


Thus far, research has largely focused on getting a deep understanding of different aspects of diversity and inclusion in OSS and contributors' experiences \cite{guizani2021long, rodriguez-perez2021slr, steinmacher2013newcomers, steinmacher2015social, trinkenreich2020hidden, feng2023state, trinkenreich2021pots}. 
It is now time to provide the aforementioned systematic monitoring support.
%
So far monitoring has been limited to research-driven dashboards focusing on the health of OSS teams and project sustainability~\cite{qiu2023climate,guizani2022attracting, ramchandran2022exploring}. 
These dashboards, however, do not look at diversity specifically. Not having such a lens can hide problem spots (such as women leaving the project) within the larger trends of overall contributors. 
In this paper, we close this gap and complement the current body of interventions by designing and deploying a systematic monitoring support focused on turnover, gender and affiliation diversity.




We developed a dynamic real-time dashboard named CommunityTapestry through participatory design (PD) \cite{schuler1993participatory}. The goal of CommunityTapestry is to: (1) signal to project leaders the specific state of their project diversity and turnover rate, (2) give them information to take actions relevant to their project needs and (3) help monitor the effects of their actions over time.

We used a large Apache Source Foundation (ASF) project, Project~B\footnote{As per our agreement with Project~B, we use pseudonyms to avoid any negative impressions about the project, as well as to protect the identity of individuals.} in our PD study. 
PD allowed us to work closely with the project community stakeholders to better understand their needs, constraints, and priorities. An additional benefit is that having the stakeholders as partners engendered an interest and investment in the project, which in turn improved the chances of the dashboard being adopted by the community. The stakeholders were the Project Management Committee (PMC) chair of Project~B, Project~B community manager, the Diversity and Inclusion vice-president of the ASF, and an engineer working on GrimoireLab \cite{duenas2021grimoirelab}, our dashboard infrastructure. We refer to them from now on as our PD partners.

Through discussions with our PD partners, we prioritized the diversity lenses (gender and organization affiliation) and the project-related activities (e.g., trends of newcomers joining or contributors leaving, communication network) of interest. We then used PD principles~\cite{schuler1993participatory} of collaborative prototyping to design the dashboard, which we implemented and deployed on existing OSS infrastructure that the project uses, and evaluated it with future users. The study spanned 14 months, during which, we created CommunityTapestry and evaluated it for inclusivity (using the \textsc{Why/Where/Fix} approach, Section \ref{sec:evaluation1}) and its usefulness 
(15 participants, Section \ref{sec:evaluation2}). 

Our results show that CommunityTapestry enabled participants to gain insight and awareness into their project diversity and turnover rate, and provided ammunition to take action.

\section{Background and Related Work}
\label{sec:background}

\textbf{Diversity lenses and turnover.}
\citet{rodriguez-perez2021slr} defined perceived diversity as innate individual diversity factors and highlights the importance of diverse Software Engineering teams. Gender diversity has received considerable attention, with studies revealing that women enhance productivity, performance, and efficiency~ \cite{rodriguez-perez2021slr, bosu2019diversity, robles2016women, ortu2017diverse, lin2016recognizing, gila2014impact, izquierdo2018openstack, robles2014floss}. However, gender biases persist in both OSS and the industry. \citet{rodriguez-perez2021slr} also emphasized the need to investigate broader aspects beyond gender diversity, such as race, age, and disability, for future research. 

With the changing OSS landscape, OSS has evolved from a volunteer-based community to a hybrid environment where company-affiliated and volunteer contributors coexist. The popularity of OSS-related business models ~\cite{munga2009adoption, spijkerman2018open, li2019does, mouakhar2017open, deodhar2012strategies} resulted in companies being involved in OSS development now more than ever. 
Contributors' diverse affiliations are vital for OSS projects, as coopetition motivates companies to join, and some donate projects to OSS foundations for neutral governance and increased contributors \cite{guizani2023rules}.  \citet{yang2022projects} found that company-based projects (59.9\% of all projects) dominate the projects joining the Apache Software Foundation (ASF), with community building and diversity as key motivators. \citet{zhang2019companies} discovered that developers from companies contribute more than volunteers in the OpenStack Foundation, but this can lead to a Pareto-like phenomenon affecting an OSS project's sustainability if dominant companies withdraw~\cite{yin2021sustainability}. 

Another important dimension of diversity is seniority. This becomes particularly significant given the overarching challenge in OSS, attracting and retaining contributors \cite{feng2022case, guizani2022perceptions}. Steinmacher and colleagues~\cite{steinmacher2015social,Steinmacher.Chaves.ea_2014} identified 58 barriers faced by newcomers and analyzed how the answers to newcomers' first emails influenced their retention \cite{steinmacher2013newcomers}.  Past works have found that most newcomers (as high as 80\% in some projects) do not become long-term contributors~\cite{steinmacher2013newcomers}. Researchers have also reported the high rate of turnover in OSS \cite{ferreira2020turnover, miller2019people, sharma2012examining} and its negative influence on team cognition and performance~\cite {levine2004,levine2005}.

\textbf{Existing interventions enhancing OSS Community Health.} 
\label{variables}
Despite the vast body of empirical knowledge on the lack of diversity in OSS and the challenges in attracting and retaining contributors, there is a lack of diversity signals for project leaders to observe in real-time the state and progress of their community. For instance,  \citet{bosu2019diversity} highlighted the importance of diversity and inclusion in OSS and the need for project leaders to appreciate and act on diversity signals. In addition, research conducted by the Linux community \cite{linuxFoundationDAndI} investigated the challenges and opportunities of fostering equity within OSS ecosystems. 



OSS community has seen the development of various interventions aimed at enhancing community health, streamlining the onboarding process, and increasing the inclusiveness of the tool itself (e.g., OSS projects). These interventions help maintainers understand their community composition and activity. \citet{steinmacher2016overcoming} introduced a portal to facilitate newcomers' integration into OSS projects, streamline their orientation, and simplify the contribution process. \citet{qiu2023climate} created a dashboard that aids maintainers in understanding their community composition and activity, including the conversation tone and overall health, and  \citet{guizani2022attracting} designed a dashboard to assist maintainers in tracking the involvement and activity of newcomers and help them attract and retain newcomers by highlighting project goals (e.g., a social good cause) and recognizing active newcomers (e.g., a ``rising contributor'' badge). Similarly, tools have been refined for inclusivity and cognitive diversity \cite{burnett2016gendermag, guizani2022debug,chatterjee2021aid, santos2023designing}. Moreover, \citet{ramchandran2022exploring} deployed a time-stamped dashboard to track the sustainability trajectories of nascent projects (incubator projects) within the Apache Software Foundation, identifying downturn events and fostering longer-term developer engagements. 

Despite progress, a gap persists: real-time diversity indicators are not readily available on OSS platforms. This makes it difficult to efficiently monitor diversity and take timely actions. Our work enriches the growing body of interventions, addressing contributors' diversity and turnover with the goal of fostering a more inclusive and diverse OSS community. We note that dashboards are widely known in software engineering \cite{storey2019software} for developer productivity \cite{meyer2019fitbit, biehl2007fastdash}, team performance \cite{gupta2018productivity}, and project health monitoring \cite{Community_2021, Zimmermann_Sadowski_2019_ch2}. Our work differs from these because of its focus on diversity and inclusion in OSS projects.


\section{Research Method}
\label{sec:design}

Our work is a PD study, where designers work closely with users to establish requirements and investigate how to combine existing technologies and structures with future technologies \cite{bodker2022participatory}. We adopted the following PD practices: (1) \textit{collaborative prototyping,} 
where non-designers can express ideas through accessible design materials and future users can gain hands-on experience with the future technology, (2) \textit{infrastructuring}, where designers and users establish organizational and technical infrastructures to ensure results of the project can be sustained after the project ends, and (3) \textit{evaluation}, where future users assess the qualities
of a design product and the outcomes emerging from the design process.

We worked with the Apache Software Foundation (ASF) and one of their projects (Project~B), a large Big-Data-related project comprising more than 1000 current contributors. The Project Management Committee (PMC) chair of Project~B, a community manager of Project~B, the Diversity and Inclusion vice-president of the ASF, and an engineer working on GrimoireLab \cite{duenas2021grimoirelab}, upon which we implement the dashboard, were our partners in the PD study (We refer to them here on as our PD partners). Through this project, we designed, infrastructured, and evaluated the CommunityTapestry over a \textit{period of 14 months}.

\subsection{Collaborative Prototyping}
\label{designsubsection}

Prototyping in Participatory Design is collaborative, where non-designers can express ideas through accessible design materials. We started the design phase by first discussing: (1) the different aspects of diversity---gender, affiliation, race, and age, and (2) the data to use in the dashboard (outcome variables)---Number of contributors left/ might be leaving, Time to merge PR.

We started by discussing the survey responses of a contributor survey that the ASF had recently completed ($N>400$), which included contributors' demographics and the challenges they faced disaggregated by demographics. We then prioritized the diversity lenses of interest to Project~B and the ASF, which could be implemented from data that GrimoireLab collects. For instance, we discussed but did not include race, an important diversity lens. Inferring race from GitHub profiles is not feasible; the definition of race is not consistent across countries and is a deeply sensitive aspect, which, if surfaced, can have a negative impact on the individual. \textit{Organization affiliation} and \textit{Gender} were selected as our lenses.

Next, we prioritized the outcome variables that were relevant to our PD partners (see Section \ref{sec:walkthrough}). For instance, we dropped the leadership outcome variables as project leadership teams tend to be small groups of people who already know each other. 

We then created interactive mockups of the dashboards using Figma\footnote{\url{https://www.figma.com/prototyping/}}. We met bi-weekly with our PD partners to refine our prototypes. During these (online) meetings, our PD partners performed walkthroughs of the dashboard while thinking aloud and sharing their screens. They provided feedback verbally or annotated on the screen share. All meetings were recorded (screen and audio).

The feedback ranged from changes or enhancement of existing features to requesting new features.  
For instance, we adjusted the network visualizations. The PR communication network, depicting communication among contributors during PR reviews (see Figure \ref{fig:dashboards-b}), was updated so that the thickness of edges between the contributor (nodes) reflected the number of PRs instead of the number of comments.
We also updated the PR communication network with a table displaying the PR details (e.g., URL, date of creation) that have not received any comments (nodes that isolate in the network visualization). 
After six meetings, the prototype features matched our PD partners' needs and constraints.

\subsection{Infrastructuring}
\label{infra}

\begin{figure}[!tbp]
\centering
\includegraphics[width=3 in]{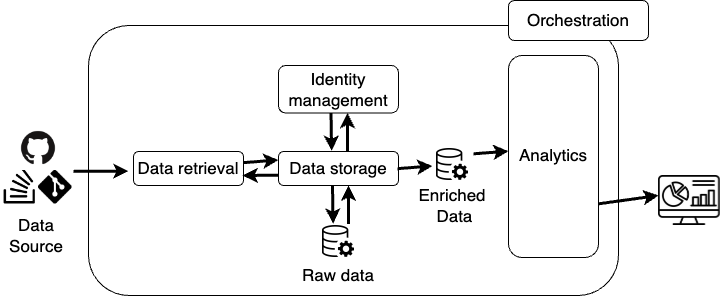}
\caption{Dashboard architecture using GrimoireLab \cite{duenas2021grimoirelab}.}
\label{fig:DashboardArchitecture}
\end{figure}

A core concern in PD is to ensure that the participating organization is in a position where experiences can be used beyond the project. We, therefore, implemented CommunityTapestry as real-time dashboards in GrimoireLab \cite{duenas2021grimoirelab}, which is an open-source analytics toolkit and a part of the CHAOSS project \cite{duenas2021grimoirelab}.
GrimoireLab allowed us to integrate real-time data across multiple platforms such as GitHub repositories and Stack Overflow (see Figure \ref{fig:DashboardArchitecture}).

\textbf{Data Mining.} In accordance with the design agreed upon with our PD partners, we used GitHub, Git, and Stack Overflow as the primary data sources for CommunityTapestry (see Figure \ref{fig:DashboardArchitecture}). GrimoireLab leverages a Python API for fetching data from repositories, which enables us to access all retrieved items from the repositories as dictionaries (JSON documents), including PRs and PR comments. The Perceval library \cite{perceval} then streamlines the process of obtaining 24 hours of continuous updates from the repository, capturing only incremental changes. This not only ensures that our dashboard stays consistently current but it also provides immediate insight into the project's ongoing activities, thereby enabling project maintainers to take proactive measures in a timely manner.

\textbf{Meta-data and identity management.} Our dashboard uses SortingHat, a relational database-powered tool, to identify the activities of human contributors in an OSS project. Bots have become increasingly prevalent in contemporary software development for tasks such as automated code review \cite{wessel2021investigating, wessel2022effects}, and SortingHat allows us to filter out bot activities.

SortingHat also helps us associate human contributors with their affiliations. For example, if a contributor's accounts in platforms such as GitHub, Git, and Stack Overflow are registered with a non-corporate domain like gmail.com, we classify their affiliation as ``affiliation unknown". However, if any of their accounts are associated with corporate domains such as google.com or apple.com, we can appropriately assign their contributions to the respective companies. This prioritization was a modification requested by our PD partners. Additionally, user IDs that are common across platforms are merged into a single entity (e.g, user@gmail and user@google.com are merged).

The data retrieved from multiple sources is then analyzed and stored in Elasticsearch NoSQL database indexes \cite{elasticsearch}. 
GrimoireELK, another component of GrimoireLab, creates enriched indexes, an abstraction customized for our visualizations, and incorporates the merged identities generated by SortingHat. The enriched indexes then feed into the Kibiter/Kibana component \cite{kibana} to generate the visualizations for the dashboard.

\textbf{Gender classification.} To infer the gender of contributors in our dataset, we used the Namsor API \cite{sebo2021performance}, a name recognition tool that estimates the gender of a full name on a -1 to +1 probability scale based on geographic information. During our initial iteration of gender classification, it became evident that relying solely on the location reported in a GitHub profile for predicting gender can lead to decreased precision. This is particularly true when the contributors' reported GitHub location does not align with the geographic origin of their name. For instance, this method posed significant difficulties when dealing with names from Asian countries such as China, Korea, and Japan. The process of transliterating names from pictographic characters to Latin letters can often lead to inaccuracies or multiple valid representations, as discussed by~\cite{kirnosova2021chinese}. As noted by Qiu et al.~\cite{Qiu.ea.2019}, this can cause the accuracy of gender prediction to drop significantly, in some cases falling below 30\%.

To mitigate this issue, we implemented a two-step process. First, we used the Namsor API to predict the origin of contributors’ names. Then, we used both the predicted origin and their full names to predict their gender. To minimize prediction errors, we applied a threshold and filtered out gender predictions with probabilities lower than 90\%  in accordance with prior literature \cite{carsenat2019inferring, sebo2021performance}.

Next, we manually enhanced the precision of gender prediction. This task was undertaken by two researchers on our team: one of Chinese origin and the other an American of Japanese descent. Together, they identified the Asian names within our dataset and then equally divided the task of manual verification.
The manual verification process entailed cross-referencing the contributors' GitHub profiles and LinkedIn pages. If a profile page contained a clear photo or a stated pronoun, we used that information. In instances where the photo was unclear, the gender was marked as unknown. It is important to note that we acknowledge gender identity extends beyond a binary classification. As such, we have designed our system to be flexible, allowing for future updates and modifications to the database.

In total, we manually validated 991 names, out of which 559 remained unpredictable, as they did not provide any profile photo on their GitHub or LinkedIn page. 191 underwent double-validation and 45 required corrections from Namsor prediction results (3 from male to female and 42 from female to male).

\section{Community Tapestry}
\label{sec:walkthrough}

\begin{figure*}[!h]
    \centering    \includegraphics[width=0.6\linewidth]{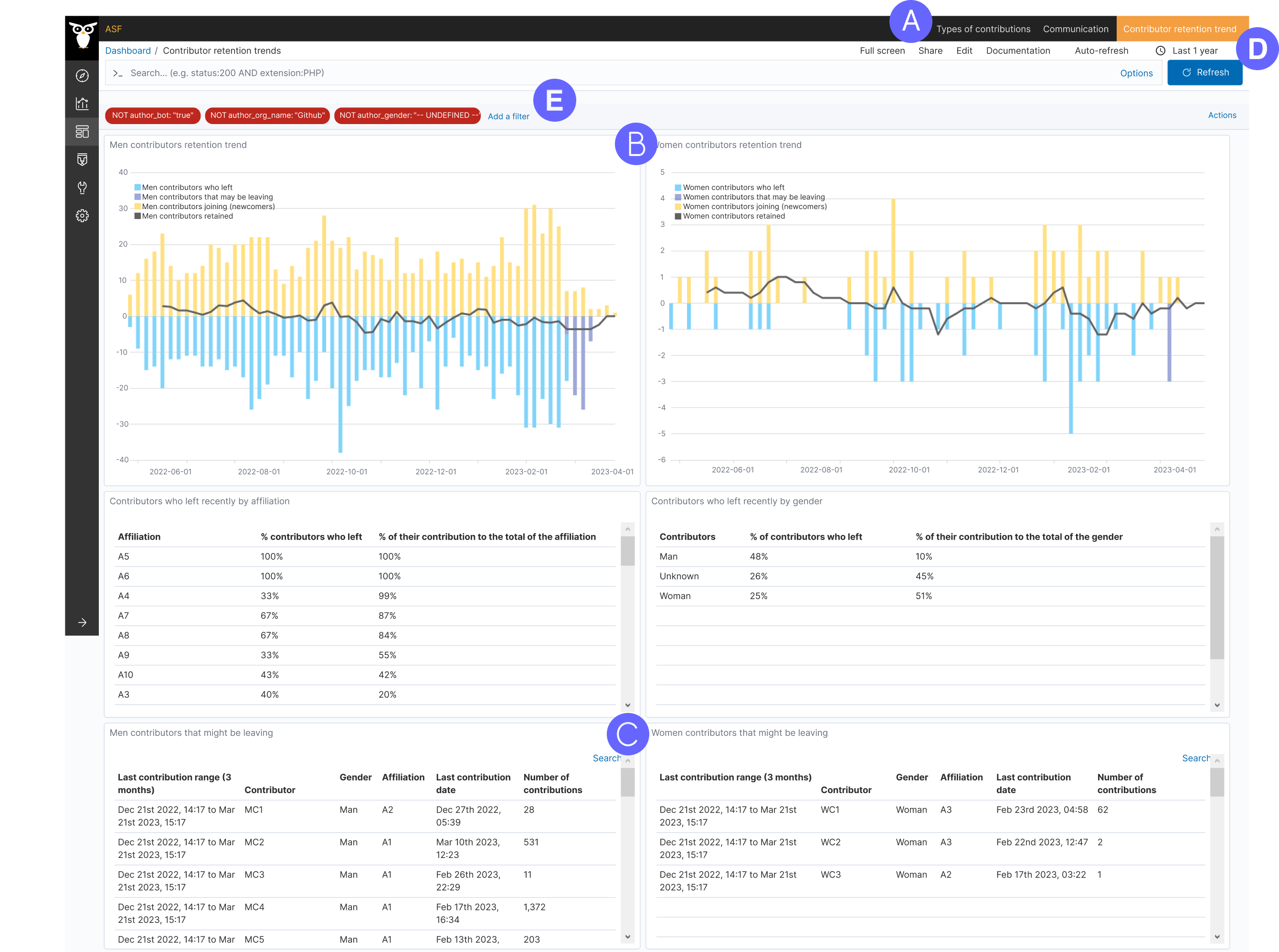}
    \caption{Walkthrough of the CommunityTapestry contributor retention trend. Contributors' names and affiliations have been deducted for anonymization purposes.``A'' stands for affiliation and ``WC'' stands for woman contributor and ``MC'' is man contributor.}
    \label{fig:walk}
\end{figure*}

\begin{figure*}[!h]
    \centering
    \subfloat[\centering Types of contributions by affiliation dashboard]{{\includegraphics[width=5cm]{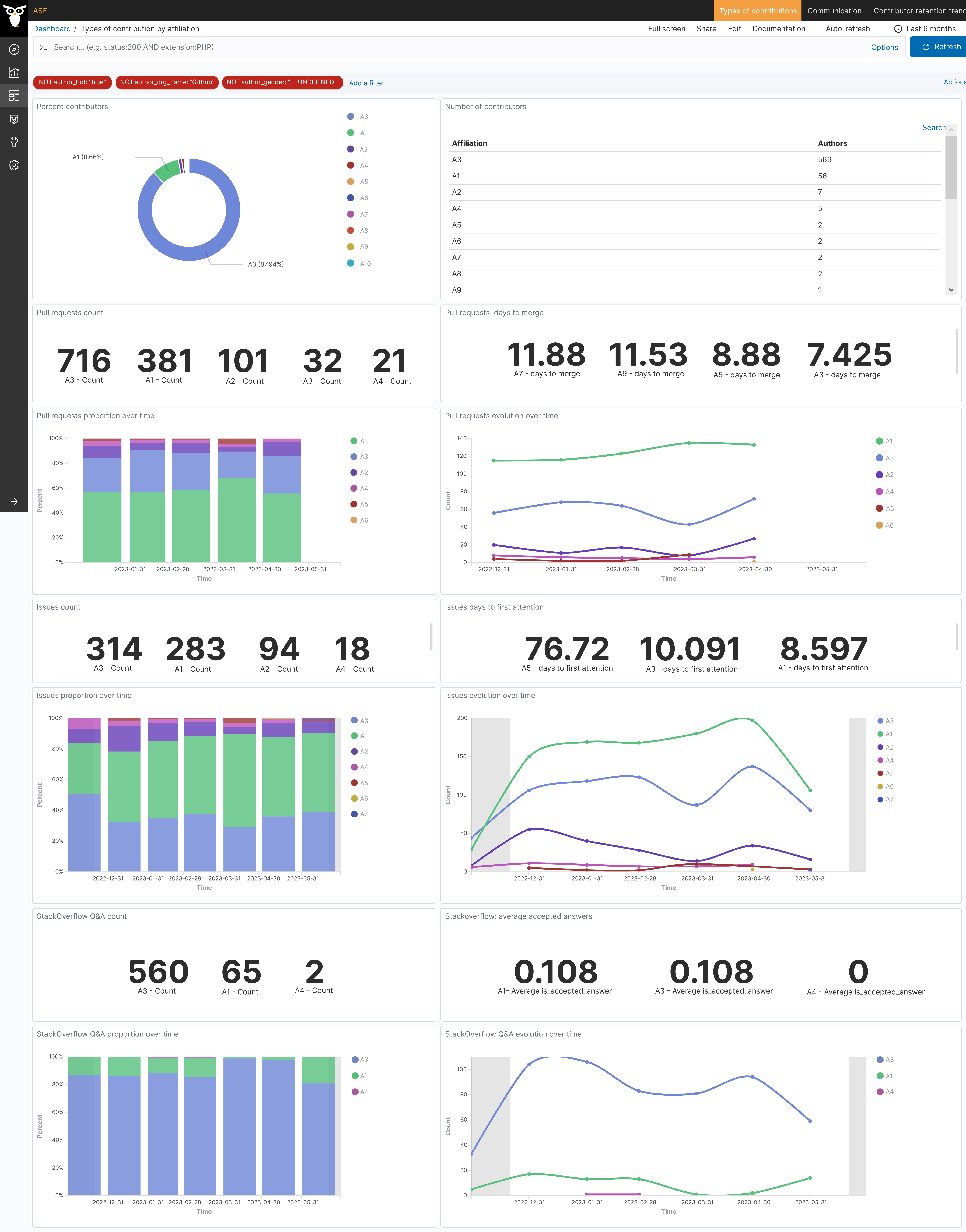} } \label{fig:dashboards-a}}%
    \qquad
    \subfloat[\centering Communication by gender dashboard]{{\includegraphics[width=9cm]{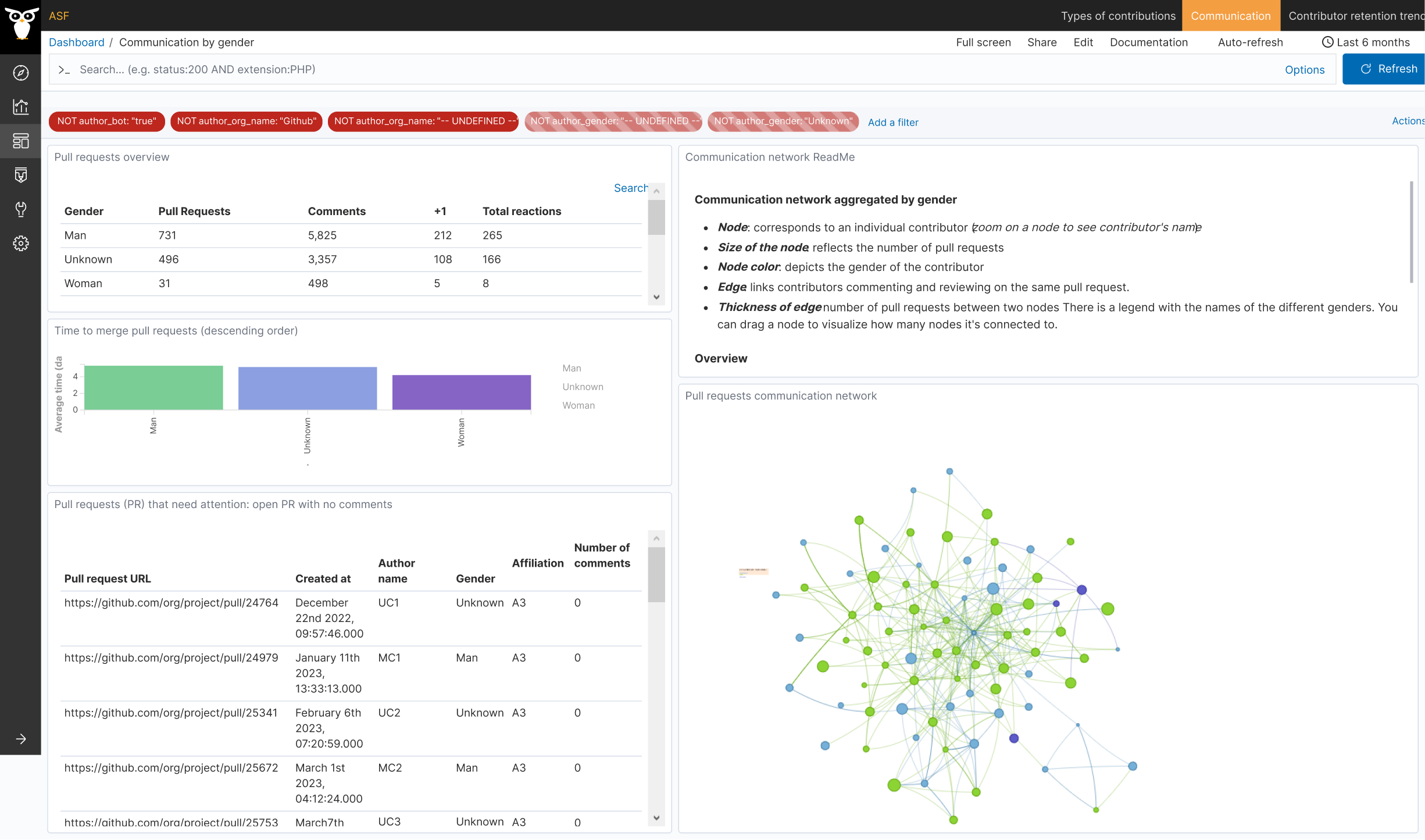} } \label{fig:dashboards-b}}%
    \caption{Example of two of the dashboards of CommunityTapestry. Contributors' names and affiliations have been deducted for anonymization purposes. See Figure \ref{fig:walk} caption.}%
    \label{fig:dashboards}%
\end{figure*}

Let us consider a hypothetical scenario in which Riley, a new PMC chair in Project~B, believes it important to be aware of and improve gender and affiliation diversity within their project. Figure~\ref{fig:walk} presents one of the dashboard pages of CommunityTapestry (see Figure \ref{fig:dashboards} for examples of other dashboards pages). The letters in the figure (e.g.,\protect\overviewLabelBig{C}.) identify the information we will explain below.

Riley begins by navigating to the dashboard ``Contribution Retention Trends'' in CommunityTapestry \protect\overviewLabelBig{A}. She notices the visualization that shows the trends of newcomers, contributors who left (i.e., inactive in the last six months), and might be 
leaving (i.e., inactive in the last 3 months), as well as the retention trend disaggregated by gender.
\protect\overviewLabelBig{B}. 
Using this information, Riley notices that while there is similar retention between men and women in the project (black line under contributor retention trends in \protect\overviewLabelBig{B}), there is a difference of 20/5 between men and women trends.

Riley decides to dive deeper. She filters the data to the last year. She reviews the detailed list of women contributors who might be leaving (the list contains name, \# contributions, affiliation, and last contribution date) to reach out to them, understands what is happening, and help if possible \protect\overviewLabelBig{C}.

Note, CommunityTapestry allows filtering the visualizations by different criteria: time \protect\overviewLabelBig{D}, affiliation name and gender\protect\overviewLabelBig{E}. These filters can be applied by clicking on the visualization or typing.

Riley wonders if the inactive women are because they were affiliated with a company that stopped contributing. She then looks at the list of contributors who might be leaving  \protect\overviewLabelBig{C}. 
Riley notices two women from the company ``A3''\footnote{We have anonymized the names and affiliations here for confidentiality reasons.} leaving the project, so she reaches out to them to find out if the reason was a personal decision or a company-related decision.


\subsection{Additional dashboards \label{sec:dashboard}}

CommunityTrapestry has two additional dashboards that can be disaggregated by gender or affiliation. Each dashboard contains different visualizations that are described below.

\textit{\textbf{Communication.}} The communication dashboard (Figure~\ref{fig:dashboards-a}) depicts the PRs interactions between contributors, disaggregated by affiliation or gender. This dashboard has four visualizations. \textit{(1) PRs overview}, which shows the number of PRs by group (i.e., gender, affiliation), comments, and likes. \textit{(2) Time to merge PRs}, displays the average time (in days) a group (i.e., gender, affiliation) has to wait to get their PR merged in descending order. \textit{(3) PRs communication network} is a graph where nodes are contributors, and the edges link contributors commenting and reviewing on the same PR. The size of the nodes reflects the number of PRs authored by a contributor, and the thickness of an edge reflects the number of PRs between two nodes (i.e., contributors). The colors differentiate the disaggregated groups (i.e., gender, affiliation) and are displayed in a legend. \textit{(4) PRs that need attention}  lists PRs that did not receive any comments and their details. This visualization contains the links to the PRs and the contributor information (name, affiliation, gender).

\textit{\textbf{Types of Contributions}}. The types of contributions dashboard (Figure~\ref{fig:dashboards-b}) details the different kinds of contributions in an OSS project using the gender and affiliation lenses. This dashboard has four sets of visualizations. \textit{(1) Contributors}, this set contains two 
visualizations presenting the percentage of contributors and the total number of contributors, both broken down by affiliation or gender. \textit{(2) PRs}, this set has four visualizations: [i] PRs proportion over time which shows the percentage of PRs during a certain period of time disaggregated by gender or affiliation. [ii] PRs evolution over time displays the evolution of PRs counts over time by group (i.e., gender, affiliation). [iii] PRs count shows the total count of PRs within a group (i.e., gender, affiliation). [iv] PRs days to merge displays the number of days before merging a PR disaggregated by gender or affiliation. \textit{(3) Issues}, this set consists of four visualizations that are similar to the ones mentioned in \textit{(2)} but present information related to the issue's contributions. \textit{(4) StackOverflow}, this last set of four visualizations contains information on StackOverflow's questions and answers. This set is similar to information in \textit{(2)} and \textit{(3)}. All dashboard pages allow filtering by different criteria such as time, gender, and affiliation. 

\section{Evaluating for Inclusivity}
\label{sec:evaluation1}

Given the focus of CommunityTapestry on inclusion, we paid special attention to ensuring that the CommunityTapestry itself is inclusive, i.e., supporting  diverse cognitive styles.
To identify and fix usability and inclusivity bugs, we used the \textsc{Why/Where/Fix} debugging process \cite{guizani2022debug}. This process is based on the GenderMag cognitive walkthrough method \cite{burnett2016gendermag} and Information Architecture \cite{informationarchitecturefortheworldwide}.
In a \textsc{Why/Where/Fix} evaluation, designers of software identify use cases for their system and perform a cognitive walkthrough to identify inclusivity bugs---why they arise (which cognitive styles are unsupported), where they manifest, and how to fix them.

The \textsc{Why/Where/Fix} process  
relies on personas with a customizable background to reflect the background of users along with a set of five cognitive styles.
During the \textsc{Why/Where/Fix} process, the participants are requested to reflect on (in)ability of the persona to accomplish specific tasks using the software, e.g., to find information about the affiliation diversity of the persona's project.

We selected the Abi persona as the cognitive styles embedded in this persona tend to be overlooked~\cite{burnett-fieldstudy-2016, burnett2017gender}. Abi-like individuals have the following cognitive styles: (i) \textit{Task-oriented motivation:} use technology for what they can accomplish with it, and not for the enjoyment of technology per se~\cite{burnett2010gender, margolis2002unlocking, burnett2011gender}; (ii) \textit{Comprehensive information processing styles:} gather fairly complete information about a task before proceeding~\cite{riedl2010there, meyers2015revisiting}; (iii) \textit{Lower computer self-efficacy} as compared to their peers; Computer self-efficacy relates with a person's confidence about succeeding at a specific task, which influences their use of cognitive strategies, persistence, and strategies for coping with obstacles; (iv) \textit{Higher risk aversion:} when trying out new features ~\cite{dohmen2011individual, charness2012strong}, which impact their decisions about the feature they use; and (v) \textit{Learning by Process:} instead of playfully experimenting (``tinker'') with software features new to them~\cite{burnett2010gender, beckwith2006tinkering, cao2010debugging}. 
We customized the Abi persona to reflect the background of a PMC member in the project (e.g., work on this particular project).

We selected five use cases that resulted in 41 evaluation questions related to users' goals and interface actions. These questions spanned each of the dashboards in CommunityTapestry: 20 questions related to the types-of-contribution dashboard, 14 questions regarding the communication dashboard, and 9 questions related the trends dashboard. Our evaluations identified a total of 10 usability bugs (24\%), 9 of which were inclusivity bugs (22\%), that is, bugs resulting because Abi's cognitive styles were unsupported. 

We then designed and implemented fixes to CommunityTaspesty. An experienced researcher with the GenderMag method then reevaluated the fixes using GenderMag moments \cite{hilderbrand2020engineering}, a fragment of a GenderMag session where the evaluation targets specific feature just-in-time. Our redesign reduced the number of usability bugs from 24\% to 5\%, and inclusivity bugs from 22\% to 2\%. Note, that the two usability bugs and the one inclusivity bug that we did not fix were related to the UI of the toolkit that we used. For example a usability bug that we could not fix is the size and position of the clock used to select the time period to display information.

\section{Evaluating with future Users}
\label{sec:evaluation2}
A key aspect of evaluating prototypes and design through participatory design (PD) is to evaluate them with future users in their natural environment~\cite{bodker2022participatory, bossen2016evaluation}. These evaluations range from informal observation, interviews to surveys. We evaluated CommunityTaspesty with \textit{future users}--Project Management Committee (PMC) members and committers of Project~B--with data extracted from their \textit{own project} through \textit{observations} of their unguided explorations of the dashboards. 

In this evaluation, we aim to answer two evaluation questions: \textsc{\textbf{EQ1:}} how the dashboards affect participants' awareness of diversity and turnover and their plans to take action? and \textsc{\textbf{EQ2}} how participants use the diversity and turnover information?

We designed our evaluation collaboratively with our PD partners. The first and third authors: (1) iterated over the study questions and tasks, and held weekly meetings with the research team to refine the study, (2) discussed the study design with the PMC Chair of Project-B, (3) sandboxed the protocol with researchers at our university until there were no more changes required to the study protocol (7 rounds), and (4) piloted the study with the PMC chair.

\boldification{**we evaluated it with these people}
\textbf{Recruitment.} Our PD partners suggested evaluating the dashboard with both PMC members and committers, since they would use the dashboard differently: PMCs will benefit from a project overview to set community goals, committers can take immediate actions such as reviewing or accepting a ``languishing'' PR. Both roles are  
important to create a healthy open-source ecosystem.

\boldification{we recruited them this way and ended up with 2 modes - zoom and in person}
We recruited participants in two stages. In the \textit{first stage}, the PMC chair shared our study advertisement with the project committee (PMC and committers) through the project's mailing list, after which we sent personalized emails to individuals in the organization whose email address was publicly available (with consent from the PMC Chair). Six participants responded, agreeing to participate (a 5.2\% response rate). These evaluation sessions were 1-hour long and conducted online (via Zoom). 
To recruit additional participants, the PMC chair recommended we attend the conference held by the organization. In this \textit{second stage}, we performed nine, in-person evaluations at the conference. 

\boldification{this is the study design, note because of difference in modality, in person had video}

\textbf{Study Protocol}
The study included 3 parts (see Supplemental document~\cite{suppDocICSE24}) and was approved by our university IRB. The first part was a pre-study questionnaire that collected demographic information about participants' gender, role, and their activities in the project. We collected this information to investigate if these aspects impact  participant interactions with the dashboards.

Next, we wanted to familiarize the participants with the dashboard features. In the online sessions, we provided participants with a link and credentials to CommunityTapestry. We then guided their exploration through the dashboard (same as in Figures~\ref{fig:walk} and \ref{fig:dashboards}, but with real project data)  with a set of question prompts such as:
``Which affiliation has had on average the longest time to have their PR merged?'', ``Have issue contributions decreased or increased among men? Among women? '' 
For the in-person sessions, since time was limited, we provided participants with short videos of the first author navigating through the same dashboard and answering the same questions as above. 

We then asked the participants to freely explore each dashboard while thinking aloud.
After they finished exploring a dashboard, participants answered Likert-scale questions on their likelihood of using the information presented in that dashboard. 
They also answered 2 open-ended questions on: (1) Is there any other way you would use the [dashboard] that we did not cover? and (2) Is there any other information you would like to see in this dashboard? For each question, participants were asked to explain their answers. 

We wrapped up the evaluation session with a post-study questionnaire, which included the questions from the pre-study questionnaire as well as questions related to the dashboard's usefulness and usability. Upon completion of the study, we thanked our participants and compensated them with a \$50 gift card as a token of appreciation, a common practice in similar studies~\cite{pater2021stadardizing}.

\boldification{Here we talk about the PMC participants}


\textbf{Participant characteristics.} 
We reached out to 23 PMC members and 68 committers. The demographics of our participants are skewed towards men (14 identified as men and only one as women as per their demographic questionnaire response). This imbalance is a reflection of the gender imbalance in the community itself. There are no women in the PMC and just seven committers who are women. The percentage of women in our participant pool matches that of the community (both at 7\%). 
Five of our 15 participants were PMC members, and ten were project committers. Their experience in their current role ranged from less than 1 year of experience to 6 to 10 years of experience. For 10 of our 15 participants, contributing to project~B was part of their primary employment. 

\textbf{Data analysis.}
\label{codes}
We used descriptive statistics and qualitative coding~\cite{gibbs2007analyzing} to analyze our data. Two researchers independently analyzed and open-coded the screen recordings and transcripts of participants from the online sessions. We used the screen recording to analyze the specific dashboard information that was used by participants and for what purpose. Then, the two researchers performed two rounds of negotiated agreement on the top-level categories (three), they then added subcategories to these resulting in a final set of eight codes. After this, one of the researchers went back to the transcripts and re-coded them as needed. The two researchers then used the resulting codebook to qualitatively code the in-person data. To ensure the reliability of the coding, they first independently coded 20\% of the data and calculated the IRR percent of agreement. The consensus level was 91\%. Given this level of consensus, the researchers split up the coding for the remainding data.

The codebook consists of three top-level categories namely \textsc{insight}, \textsc{explain}, and \textsc{action}, that reflects how participants verbalized the use of the information they found in the dashboards in their free-form exploration. For example, when participants verbalized synthesized information that they have formed from the dashboard (e.g., \textit{``I am surprised by this one, [name of company] because I don't know it... ''(P3)}), we coded such information as an \textsc{insight}. 
In cases where the participant shared a reason or explanation (e.g., \textit{``...[company name] is consulting company. So they will do some projects, and you know they'll do project end to end...and they will kind of loop out...'' (P1)}), we coded it as an \textsc{explain}. 
In some cases, CommunityTapestry served as a springboard for participants to take and/or think of future actions (e.g., \textit{``...almost certainly that would be a good time to reach out to somebody because that's somebody who is given a lot to the project and understanding what is going on...''(P4)}) to which we assigned the code \textsc{action}. 

These three top-level categories were then enriched with 8 sub-categories. First, \textsc{insight} and \textsc{explain} were subdivided into \textsc{project} and  \textsc{case} related. For instance, an explanation could be about a project (e..g, its culture or practices) or about a specific case that a participant was exploring in the dashboard (e.g., a specific contributor who left the project).
Second, we subdivided \textsc{action} into three types of future actions that participants verbalized they would take (i.e., \textsc{action-community}, \textsc{action-personal}, \textsc{action-affiliation}). A fourth code, \textsc{action-explore}, was used to identify cases where participants navigated outside of the CommunityTapestry (e.g., GitHub profile) for additional information.

\subsection{Findings}
 \begin{figure}[!tbp]
    \centering
    \includegraphics[width=3.4 in]{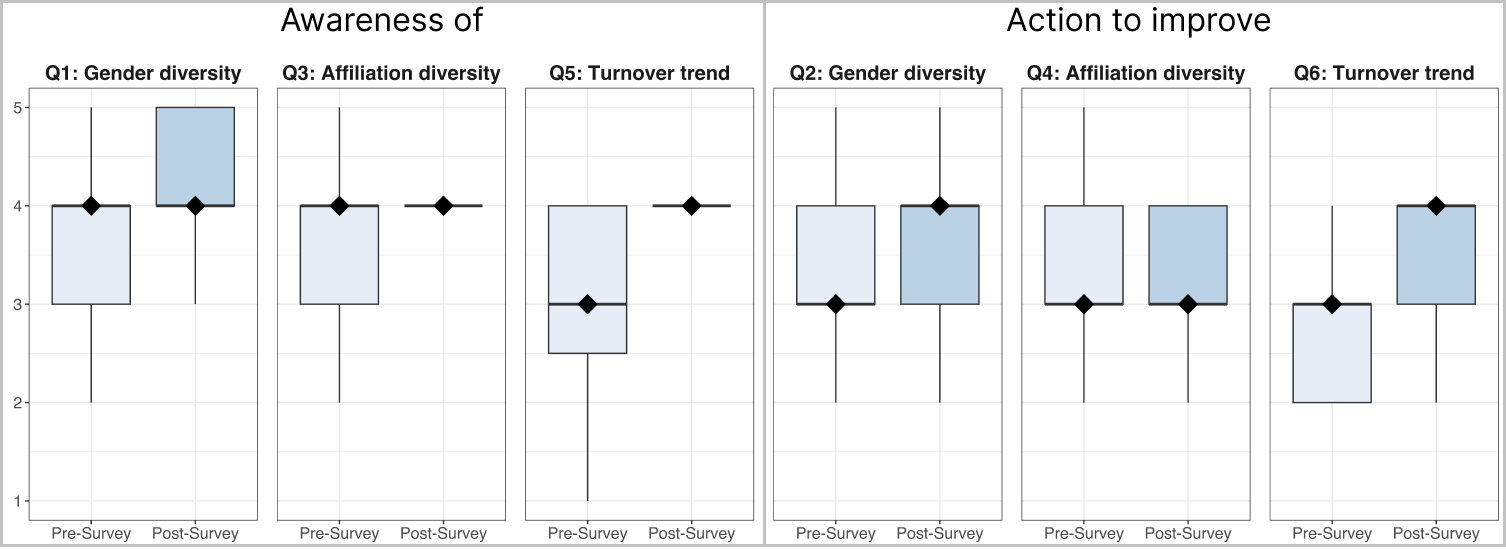}
    \caption{Likert Scale responses for Pre-Survey and Post-Survey.
    Black dots:  median responses. Outliers are not depicted to emphasize the overall trends.}
    \label{fig:boxplot}
\end{figure}
Next, we discuss the overall usefulness and diversity awareness before and after the introduction of CommunityTapestry. 
We then detail how participants used the information of CommunityTapestry and their likelihood of using this information in the future. 

 \begin{figure*}[!tbp]
 
    \centering
    \includegraphics[width=6 in]{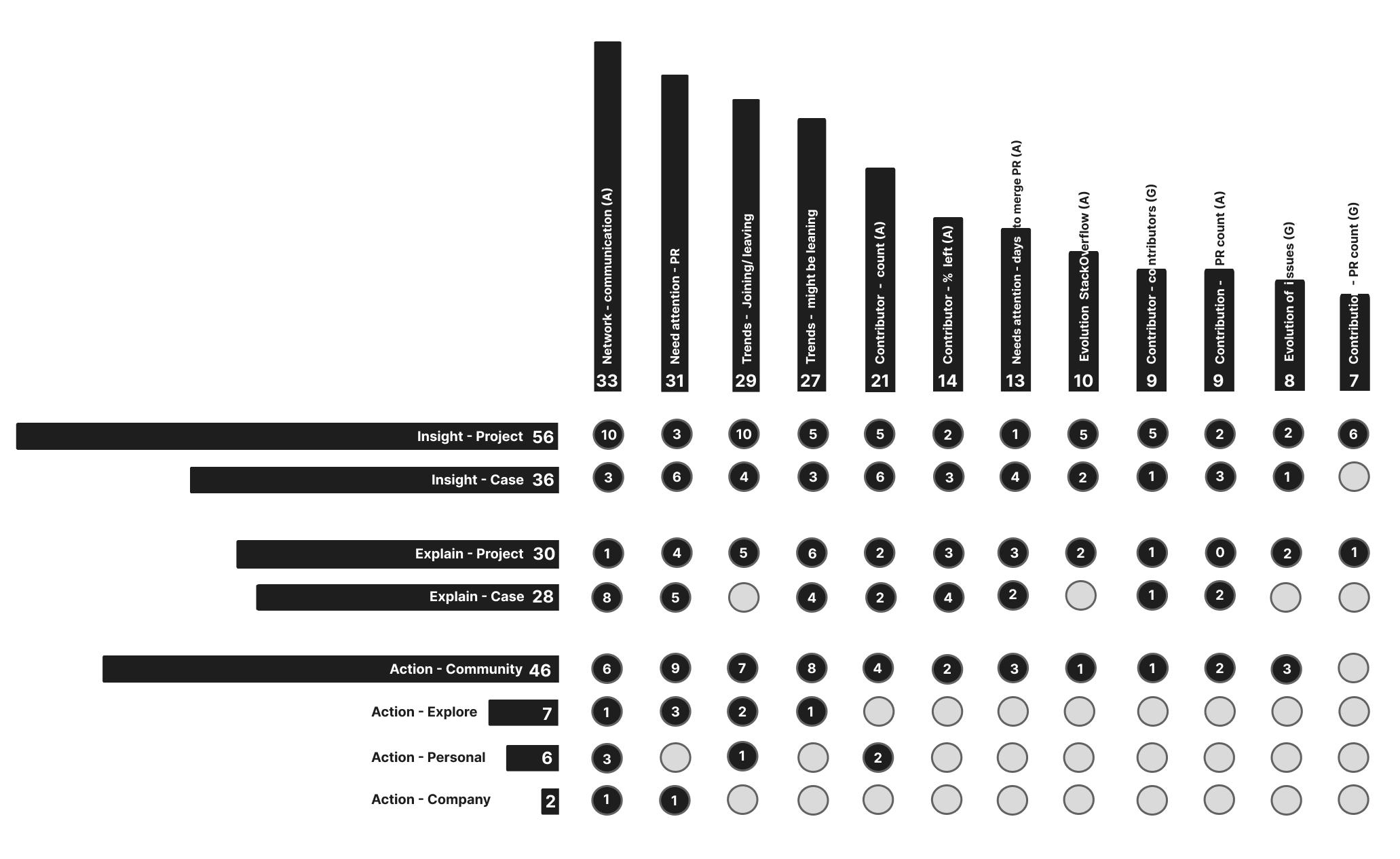}
    \caption{Information usage. The number in the circles show the number of participants who used [specific] information for \textsc{insight},  \textsc{explain}, or \textsc{action}. 
    The vertical bar charts reflect the frequency of usage of an information source and the horizontal bar charts denote the frequency of a particular information use (e.g., \textsc{insight-project}) across all participants. 
    Information sources are annotated with an (A) or (G) when that information source was disaggregated only by that lens (Affiliation or Gender). 
    }
    \label{fig:coolFigure}
    \end{figure*}

\subsubsection{EQ1: Overall usefulness and awareness.}

 We analyzed the responses from two 5-point Likert scale questions on the dashboard usefulness and plan for continued use from Qiu et al.~ \cite{qiu2023climate}. Overall, participants reported that the dashboard was useful to them (Q\_usefulness: Med = 4, SD = 1) and that they would continue to use it (Q\_continue to use: Med = 4, SD = 0.64). CommunityTapestry's potential to inspire action was reflected by the post-survey responses where most participants reported that the dashboard helped identify aspects to improve (Med = 4, SD = 0.74) and will have an effect on their actions (Med = 4, SD = 0.88). 

Figure \ref{fig:boxplot} depicts the distributions of Likert scale responses in the pre- and post-evaluation questions about their \textit{awareness} of each diversity lens (e.g., awareness of gender diversity) and \textit{plan to take action} questions (e.g., I plan to take action to improve gender diversity).  
The distribution of responses (Figure~\ref{fig:boxplot}) for awareness questions (Q1-gender, Q3-affiliation, Q5-turnover) show an upward distribution shift between the pre- and post-evaluation questions, although the median (Med=4) stays the same for Q1 and Q3. The reason for the median staying the same could be because of self-selection bias; participants who volunteered might those who care about and are already aware of the diversity in their project. There is a 1-point improvement in awareness for Q5. The increase in median awareness about turnover could be due to the unexpected nature of this information as explained by P5, \textit{``So that [turnover] definitely is generally like high across the boards for percent who left I wouldn't have expected it to be that high.''(P5)}    

When it comes to planning to take action to improve, we see distribution shift for Q6-turnover, and median shifts for Q2-gender and Q6-turnover. A reason for both distribution and median shifts for turnover could be its importance to project health, as P4 explains \textit{``if you notice trends of either a specific gender or a specific affiliation leaving that indicates that we might need to act in some way.''} 
Overall, while CommunityTapestry did not have an effect on action to improve gender and affiliation diversity, it helped move the awareness distribution upward for gender, affiliation, and turnover and participants reported they planned future action to improve the turnover trend within their project. 

\subsubsection{(EQ2): Information Use} We first look at how the information was used (\textsc{EQ2a}) during the think aloud sessions and then likelihood of participants using this information for their own use based on the post-study questionnaire (\textsc{EQ2b}). 

\textbf{\textsc{EQ2a} - Information usage.}
In their free form explorations, participants used the dashboard information to (1) get insights, (2) get an explanation for the insight, and (3) take action at different levels, as reflected in our codeset in Section~\ref{sec:evaluation2}. 
Figure \ref{fig:coolFigure} depicts how participants used the dashboard information. The first dimension (vertical bar chart) shows the frequency of insights, explanations, and actions generated from a particular piece of information. The second dimension (horizontal bar chart) shows the frequency of which information was used (i.e. insights, explanations, actions). 

Participants most frequently used CommunityTapestry to get insights about the project (Figure \ref{fig:coolFigure}, rows 1 and 2) whether at the project level or related to a specific case (e.g., a specific affiliation leaving the project). Actions in general and actions related to the community, were the second most frequent information usage. 

Next, we present specific examples of information usage, by selecting the top three information sources in Figure \ref{fig:coolFigure}: (1) PR communication network (2) PRs that need attention, and (3) overall contributors joining/ leaving.

\textit{\textbf{PR communication network}}: While navigating the dashboard, participants verbalized insights they gained about their projects. For instance, Participants (P1, P4, P6, P9, P12) realized that they had cliques because of their project review culture  \textit{``it does stand out that there are these clusters of people with same affiliations reviewing each other's code'' (P1)} and \textit{``[nodes] contributing to [project], but they are not attached to the main network'' (P12)}.
They had a possible explanation for it: \textit{``the main contributors to [Project~B] are people inside [main affiliation]'' (P12)}. 
Seven participants (P2, P3, P4, P5, P6, P7, P9, P14) articulated explanations for specific cases they observed. For example, P5 noted a possible reason for the clusters in the network: \textit{``[they are] working on maybe this at the same company on the same features, and therefore you have you know, you're communicating more because you have more context that's shared'' (P5)}. P7 explained the reason for the main cluster in the network: \textit{``[Company employees] are reviewing everything, and so pretty much all nodes are going to be connected to [company], and we've learned to review each other's code a lot'' (P7)}. P9 explained reviewer clusters were because: \textit{``the project is quite large, It is not possible to understand a piece of code for everybody'' (P9)}. 

However, the reviewer clusters made them realize that community members should aim to learn from each other: \textit{``knowledge sharing is important as well because at least key contributors ... should be aware of what is happening'' (P9)}. 

In total 6 participants (P1, P2, P3, P4, P5, P9) were inspired to take a community-related actions. For instance, some shared that they would \textit{``try to bring them [isolated sub networks] into the fold'' (P2)} and \textit{``split the [review] load that goes to this person to not block'' (P3)}.
The communication network inspired other types of actions (P1, P5, P8, P6). For example, P1, P5, P8 said they would use the network (A) as a recognition mechanism: \textit{``this dashboard will correspond to our job, or like our productivity'' (P8)} and \textit{``as an employee of a company that wants to get promoted. I might use this to show my impact to show that I am supporting lots of people doing reviews for lots of people'' (P1)}. P5 stated that they would use the network to: \textit{``know who is well connected,...If there was a really big change that I wanted to try to orchestrate...so that would be a really good person to reach...getting their thoughts on it early'' (P5)}. 



\textit{\textbf{PR that need attention}}: When reviewing the table of PRs that need attention, participants realized the large number of PRs without any comment. Participant P8 explained: \textit{``We have a PR bot to auto-assign reviewers ... Although this bot I know, I often ignore it because it's sometimes accurate, and sometimes it makes some noise'' (P8)}. Participants felt that the overall table to be \textit{``very helpful for the maintainer'' (P8)}. For individual contributors, the project context was instrumental in helping them use this information. For instance, P1 looking at a particular PR in the table explains: \textit{``he's [PR author] kind of actually a fish in the water in the community... He doesn't need help just because I know him. And so in this case, if this PR hasn't received comments, it's because he's not interested in moving it forward'' (P1)}. On the other hand, when observing the same information about another contributor, P1 shared: \textit{``out of the people that I don't know, or that I know, that are not inside the community very much, I might look at their PRs in detail'' (P1)}.

The information provided by CommunityTapestry inspired som participants (P1, P2, P3, P6, P12) to explore outside of the tool and get additional information such as GitHub PR, e.g., when participants noticed an affiliation or newcomers with a high number of contributions that they did not know (P1, P3, P2, P6), or when they noticed a PR that needed attention (P1, P3, P12)---one that did not receive any comments yet \textit{``can I check this one? I think there are comments [in the PR body in GitHub]... Oh, it is the same person commenting on himself...that doesn't count'' (P12)}. 

\textit{\textbf{Trends of contributors joining/leaving}}: Participants (P4, P5, P13) shared project insights and possible explanations about their community composition in terms of gender where \textit{``it definitely seems like statistically, there are more men that join per month the project than women'' (P13)} and the trends of joining and leaving suggest the presence of one time contributors \textit{``the charts are very correlated like for example, here the newcomer's line and the people who left line are both very long. It makes me think that there might be people that make one single contribution'' (P1)}. 

These insights inspired participants to reflect on how to increase contributors retention, especially of women contributors. For instance, participants shared the need to \textit{``connect them [newcomers] to people who have been retained previously so that they can kind of have a more personalized experience in how they can be retained'' (P2)} and \textit{``trying to be extra prompt on PR reviews and kind of flagging PR reviews from you know, under-represented groups, can be helpful'' (P4)}, especially when \textit{``the number of women is small enough that you can look at individual cases'' (P1)}. Participants recommended to involve active newcomers in further aspects of the project: \textit{``They [newcomer] are like, having have contributed very, very big things to me that I would, involve them in a design review'' (P11)}.

 \begin{figure}[!tbp]
    \centering
    \includegraphics[width=3.5 in]{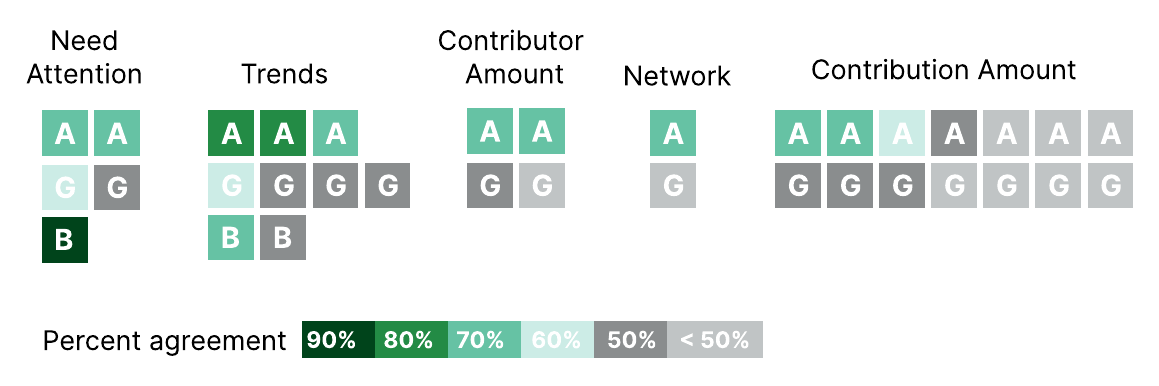}
    \caption{Dashboard information sets ranked by usage likelihood. The likelihood is a response to Likert Scale questions about participants using different pieces of information. Each square represents an information source within CommunityTapestry. ``A'' or ``G" represent information sources disaggregated by affiliation or gender, respectively. ``B'' denotes both diversity lenses. 
    }
    \label{fig:LikertInfo}
    
    \end{figure}

\textbf{\textsc{EQ2b} - Likelihood of information usage.}
After using each dashboard participants provided their opinion on the likelihood (5-point Likert scale) of using the information provided in that dashboard. Overall there were 34 information sources.
We grouped related information sources together into \textit{five information sets} (Figure \ref{fig:LikertInfo}). For example, the \textsc{Need Attention} set includes information such as, ``days to merge PR'', ``days to first attention to issues'' and ``PR that needs attention''. In Figure \ref{fig:LikertInfo} each box represents an information source and is annotated by ``A'', ``G'' or ``B'' if that information source was disaggregated by Affiliation-only, Gender-only, or both, respectively. Using participant's Likert scale responses, we calculated the percentage of participants who were at least ``likely'' to use said information (see the colors in Figure \ref{fig:LikertInfo}).


Information that participants reported they were most likely to use was in the \textsc{Need attention} set. For example, 90\% of participants were at least likely to use the ``PRs that did not receive any comments'', both for gender and affiliation disaggregations (see Figure \ref{fig:LikertInfo}, row 3, col 1). This is probably because the information in this set allows participants to \textit{take remediation action}. For instance, P15 shared: \textit{``I might have unconsciously prioritized contributions from advanced users of Project~B, and I think for the project to be successful we should increase the state of affiliation diversity'' (P15)}. Others reported that they actively sought out items that needed attention and the dashboard makes it easier. P4 said:\textit{``We already have patterns of rotating through and looking for PRs that are out of date, and this would make it a lot easier''}.

Information about \textsc{trends} was the next most likely to be used information. The trends were useful for participants to identify ``worrisome" patterns that would prompt them to take action. For example, P9 shared \textit{``these numbers, of course, are interesting to see like a strong trend that in some ways is a sign to talk to people that represent such groups''}. Another way trends were useful was to understand the impact of a change or decision. P4 said:
\textit{``overtime view is helpful for evaluating tools [migrating from Jira to Github issues]...did that have an impact on specific contributors...did that help more people from these backgrounds enter into the project'' (P4)}. 

Participants 
preferred the\textit{ affiliation lens over the gender lens} across the  information sets (Figure \ref{fig:LikertInfo}). This might be due to the fact that there is already some level of awareness about gender diversity in the project, as explained by P1  \textit{``there's not that many women, so we can just look at individuals directly we don't need to look at percentages'' (P1)}. Further, Project~B is used by multiple companies that have dedicated contributing teams where \textit{``if there was sort of an unexplained mass exodus from a given affiliation...that might have a really big impact on the overall health of the project'' (P5)}.

\section{Discussions}
\subsection{Reflections from participatory design.} 
A key principle of PD is mutual learning between users and designers and reporting reflections on the process through which the product is developed \cite{schuler1993participatory}.
To help with our learning of the organization needs, we selected our PD partners to include a wider set of stakeholders than just members of Project~B. Each of our PD partners had different priorities and concerns and our  discussions helped in  mutual learning of the different constraints governing our PD partners. For example, some of the diversity lenses of particular interest to the ASF (region, English proficiency) was of less concern to Project~B, whose context was a company sponsored project that is seeing rapid growth and adoption. At the same time, we had to align the constraints of our infrastructure provider, where a key constraint was the ability to maintain the infrastructure and work using the information sources that they were already collecting. Additional information sources such as, social media or event activities that signal non-code activities was of particular interest to Project~B's community manger, but it turned out to be infeasible to be added to GrimoireLabs, the underlying dashboard infrastructure.


As researchers, we learned several things. We better understood the conflicting priorities existing in a hybrid project that includes paid contributors and volunteers. For example, the Project~B PMC chair had to be particularly cognizant of ``satisfying the needs'' of contributors from company sponsors, before being able to put resources to meet other needs. They also had to navigate the ``politics'' of reviewer cliques and how this reflected on the project, especially to those outside the company. There was concern about whether patterns that were now visible through the dashboard would have a negative impact on morale and sponsorship, both of which impact retention. We had long conversations of what type of demographic information could be inferred and if they should be surfaced in the dashboard and whether the dashboard should be public. At the end, it was decided to have the dashboard visible only to the PMC, then increase the visibility to those who have commit access. Whether the dashboard would be public would be determined based on how the community adopts the dashboard and its impact on the project. 

Finally, our conversations made us realize that the information extracted from the dashboard, its signals, and its fit within the future users' process was more important than the dashboard itself. Thus, we focused a large portion of our evaluation on this topic.

\subsection{Limitations}
\label{sec:limitation}

This section presents the limitations and the reliability of our results from the perspective of qualitative work ~\cite{MerriamBook} as well as quantitative analysis ~\cite{lincoln1985trustworthiness}.
%
Our decision to use Participatory Design (PD) as our research method, which contextualizes the results to a specific organizational context, affects the generalizability of our findings. It is possible that other communities with different demographic distributions may prioritize different diversity lenses and find different information sources to be more useful. However, we believe that PD allowed us to deeply understand the organizational context and needs of Project~B and situate the intervention to their needs, which outweighed PD's drawbacks. Another factor to consider is the small sample size (15) of evaluation participants. We attempted to mitigate this threat by recruiting both PMC members and committers, both from online venues as well as from the organization-sponsored conference. The evaluation of the impact of our prototype is left as the future. The next steps in the PD will examine whether Project~B adopts the CommunityTapestry and in what capacity.

Another issue could be the reliability of our findings. We sought to mitigate this issue using validated questions and questions from prior literature when possible. We worked closely with our PD partners and sandboxed and piloted the evaluation instruments to avoid any such issues. The second issue could be the reliability of our qualitative analysis. To avoid misinterpretation in the qualitative coding of our data, two researchers independently coded the data and performed two rounds of a negotiated agreement for the online evaluations (phase-1), and got 91\% IRR when coding the data from the in-person evaluations (phase-2).

For the quantitative analysis, first and foremost, we acknowledge that gender is not binary and that the only reliable way to know one's gender identity is to ask the person themselves. However, such an approach does not scale, so we had to use algorithmic tools such as NamSor. We attempted to reduce potential errors to the extent possible as explained in Section \ref{sec:design}. We stress that this is a limitation of the evaluation rather than of CommunityTapestry itself: in the future, OSS projects might adopt different ways of recording the contributors' gender, e.g., by expecting them to indicate their preferred pronouns on GitHub---the way CommunityTapestry works will not be affected.

Finally, we acknowledge the possibility of self-selection bias, where evaluation participants might have been contributors who cared about diversity in their projects. We minimized the effects of this by reaching out to participants with different levels of expertise and agency in the project (i.e., PMC members, committers). Exhaustion and distraction may also have affected online evaluation participants. To minimize this risk, each session lasted at most one hour. We also made an effort to scope our evaluation questions.

\section{Concluding Remarks}
\label{sec:conclusion}
In this paper, we reported our experiences and outcome in designing, developing, and evaluating a dynamic real-time dashboard, called CommunityTapestry. CommunityTapestry prioritizes three factors: gender diversity, organizational affiliation diversity, and contributor turnover. First, the focus on gender diversity acknowledges the challenges OSS is facing in this area. The dashboard aims to promote inclusivity and track progress within the community by monitoring project signals. Second, it emphasizes affiliation diversity, recognizing the evolving nature of open-source collaboration. By capturing the involvement of both companies and volunteer contributors, the dashboard provides insights into the changing dynamics of OSS projects. Finally, the dashboard helps monitor contributor turnover, a crucial aspect of sustaining OSS projects. Tracking turnover rates can enable projects to identify negative trends and implement solutions to retain contributors.

Our evaluation of CommunityTapestry with future users shows promise about the usefulness of aggregating  information as done in our dashboards. Evaluation participants were particularly interested in tracking trends among contributor turnover with a focus on affiliation diversity. The dashboards helped participants get insight about their project practices (e.g., reviewing culture), be able to explain certain phenomenon (e.g., an explanation of why a group of contributors became inactive), and most importantly have the information to take action. For instance, participants were interested in monitoring items that needed attention and trends of contributors who might be leaving to take remedial actions. 
where \textit{``the good thing about might be leaving is that it could be actionable, you might be able to do something about it'' (P5)}. 

The ability to drill down on specific items (e.g., contributor details of someone who hasn't received PR comments) allowed to either know what action to take (e.g., respond to PR if a newcomer) or recognize that its not a concern (e.g., a core contributor who does not need any input). 
Participants recognized the dashboards' potential not only in the short term but also in the long run, as highlighted by P7: \textit{``I think this is like the best dashboard for overall project help that I would look at and again gives me the chance to see the results of targeted interventions most clearly'' (P7)}.



\clearpage
\balance
\bibliographystyle{ACM-Reference-Format}
\bibliography{bibliography}
\end{document}